\documentstyle[11pt]{article}
\begin{document}

\title{\bf Excitation of resonators by electron beams}
\author{Yukio Shibata$^{\dagger}$, Satoshi Sasaki$^{\dagger}$,
Kimihiro Ishi$^{\dagger}$, Mikihiko Ikezawa$^{\dagger}$, \\
\small \it $^{\dagger}$ Research Institute for Scientific Measurements,
Tohoku University, Japan \\ E.G.Bessonov$^{\dagger \dagger}$\\ \it
\small  $^{\dagger \dagger}$ Lebedev Physical Institute AS, Moscow,
Russia \date{} } \maketitle

\begin{abstract}
In this paper the main consequences of the vector theory of excitation
of resonators by particle beams are presented. Some features of
excitation of broadband radiation in longitudinal modes of the enclosed
and open resonators are discussed.
\end{abstract}

\section{Introduction}

The excitation of resonators is described by Maxwell equations in
vacuum \cite{landau}  - \cite{heitler}

     $$div\vec{E}=4\pi \rho \hskip 4mm (a)\hskip 8mm
     rot\vec{H} = \frac{4\pi}c\vec{J} +\frac
     1c\frac{\partial\vec{E}}{\partial t} \hskip 4mm(b),$$
     \begin{equation} rot\vec{E} = -\frac 1c\frac{\partial
     \vec{H}}{\partial t} \hskip 4mm (c), \hskip 18mm div\vec{H}=0
     \hskip 4mm (d).  \end{equation} 

These equations are a set of two vector and two scalar equations for
vectors of electric $ \vec{E}(\vec r,t)$ and magnetic $\vec{H}(\vec
r,t)$ field strengths or eight equations for six independent components
of the electric and magnetic fields. We suppose that the charge density
$\rho (\vec r,t)$ and current density $\vec J(\vec r,t)$ are given
values. It means that only four components of the electromagnetic field
strengths are independent.

The solution of these equations includes transverse electromagnetic
field strengths of free electromagnetic waves $\vec E ^{tr}$, $\vec H
^{tr}$ and accompanied longitudinal electric field strengths $\vec E
^l$ of Coulomb fields of the beam crossing the resonator. Transverse
electromagnetic field strengths excited by the beam in the resonator
comply the condition $div \vec E ^{tr} = div \vec H ^{tr} = div \vec H
= 0$. Longitudinal electric field strength comply the condition $rot
\vec E ^l = 0$, $div \vec E ^l = 4\pi \rho$ \cite{jackson} -
\cite{lopuhin}\footnote{In general case transverse fields are not only
free electromagnetic waves.  Both a static magnetic field, a magnetic
field accompanying a homogeneously moving particle and arbitrary time
depended magnetic field are transverse one. A part of the Coulomb
electrical field accompanying a relativistic particle is transverse one.
The most simple example of the transverse electric field
strength is the electric field strength of the homogeneously moving
relativistic particle $\vec E ^{tr} = \vec E - \vec E ^{l}$, where
$\vec E = e\vec r/\gamma ^2r^{*3}$, $\vec E ^{l} = e\vec r/r^{3}$, $\vec
r$ is the radius vector directed from the particle to the observation
point, $\gamma = \sqrt{1 - \beta ^2}$ relativistic factor of the
particle, $R ^* = (x -vt)^2 + (1 - \beta ^2) (x^2 + y ^2)$, $\beta =
v/c$, $v$ the velocity of the particle \cite{landau}, \cite{jackson}.
After a particle beam cross a resonator then only transverse free
electromagnetic waves stay at the resonator.}.  Free electromagnetic
fields in resonators are solutions of homogeneous Maxwell equations
($\vec \jmath = \rho =0$) with corresponding boundary conditions. These
solutions are a sum of eigenmodes of the resonator which include a
discrete set of eigenfrequences $\omega _{\lambda}$ and corresponding
to them functions $\vec E_{\lambda}(\vec r,t), \vec H_{\lambda}(\vec
r,t)$ for the electric and magnetic field strengths (further we will
omit the superscripts $tr$ and $l$ in the fields). The subscript
$\lambda$ includes three numbers ($m,n,q$) corresponding to transverse
and longitudinal directions of the resonator axis. In the case of open
resonators the transverse electromagnetic $TEM_{mnq}$ modes are
excited.  When the number $q$ is very high then this number is omitted.
Usually in the open resonators many longitudinal modes are excited even
in the case of free-electron lasers emitting rather monochromatic
radiation.

The solution of the problem of excitation of resonators is simplified
by introduction of a transverse vector potential $\vec A(\vec r,t) =
\sum _{\lambda} \vec A_{\lambda} (\vec r,t)$ of free electromagnetic
fields in Coulomb gauge $div \vec A = 0$, where scalar potential
$\varphi = 0$ when $\rho = 0$ (here we omitted the superscripts $tr$
and $l$ in the vectors $\vec A ^{tr}$). The corresponding wave equation
for this vector can be solved by the method of separation of variables
when we suppose $\vec A_{\lambda} (\vec r,t) = q _{\lambda}(t)\cdot
\vec A_{ \lambda}(\vec r)$, where $q_{\lambda}(t)$ is the amplitude of
the vector potential and $\vec A_{\lambda}(\vec r)$ is the
eigenfunction of the resonator normalized by the condition $\int |\vec
A _{\lambda}(\vec r)|^2dV =1$. In this case the total free
electromagnetic field in the resonator is described by the expression
$\vec A(\vec r,t) = \sum _{\lambda} \vec q_{\lambda}(t) \vec
A_{\lambda}(\vec r)$.

The electric and magnetic field strengths of the transverse free fields
in resonators can be expressed through the vector potential in the form
$\vec E_{\lambda} (\vec r,t) = -d\vec A_{\lambda}(\vec r,t)/d\,ct = -
\dot q _{\lambda}(t) \cdot \vec A_{ \lambda}(\vec r)/c$, $\vec H
_{\lambda} (\vec r,t) = rot \vec A _{\lambda} (\vec r,t) = q
_{\lambda}(t)\cdot rot \vec A_{\lambda} (\vec r)$, where $\dot q
_{\lambda}(t) = d\,q _{\lambda}(t)/d\,t$. When the charge and current
densities are in the resonator then a scalar $\varphi _{\sigma}$ and a
longitudinal vector potential $\vec A ^{l}$ ($rot \vec A ^{l} = 0$)
determine Coulomb fields of the beam in the resonator. We are not
interesting them in this paper.

When active and diffractive losses in the open resonator are absent
then the vector potential of a free electromagnetic field in the
resonator excited by the beam can be presented in the form

     \begin{equation}
     \vec A(\vec r,t) = \sum _{\lambda} q_{m \lambda} \vec
     A_{\lambda}(\vec r)e^{ i\omega _{\lambda}t},
     \end{equation} 
where the coefficient $q _{m \lambda}$ is the amplitude of the excited
eigenmode.

The electromagnetic fields excited by the electromagnetic beam are
determined by the nonhomogeneous Maxwell equations or the
corresponding equation for the vector potential

     \begin{equation}
     \Delta \vec A(\vec r,t) - {1\over c ^2}{\partial ^2\vec A(\vec
     r,t)\over \partial t ^2} = -{4\pi \over c}\vec J(\vec r, t).
     \end{equation} 

The solution of the Eq(3) can be found in the form $\vec A(\vec r,t)
= \sum _{\lambda} q_{\lambda}(t) \vec A ^{tr} _{\lambda}(\vec r) +
\sum _{\sigma} q _{\sigma}(t) \vec A _{\sigma}^{l}(\vec r)$ (here we
stayed superscripts $tr$ and $l$ and used the conditions $\vec A
_{\lambda,\sigma} (\vec r,t) = q _{\lambda,\sigma}(t) \cdot \vec A
_{\lambda,\sigma}(\vec r)$). If we will substitute this expression into
equation (3), integrate over the volume of the resonator, use the
condition of normalization $\int |\vec A_{\lambda}(\vec r)| ^2dV = \int
|\vec A _{\sigma}(\vec r)| ^2dV = 1$, $\int \vec A_{\lambda}(\vec
r)\vec A_{\lambda ^{'}}(\vec r)dV = \int \vec A_{\lambda}(\vec r)\vec
A_{\sigma}(\vec r)dV = \int \vec A_{\sigma}(\vec r)\vec A_{\sigma
^{'}}(\vec r)dV = 0$ and take into account that the vector $\vec
A_{\lambda}$ comply with the condition $\Delta \vec A_{\lambda} = -
(\omega _{\lambda}/c)^2\vec A _{\lambda}$ then we will receive the
equation for change of the amplitude of the eigenmode $q _{\lambda}$
for free fields in the resonator

     \begin{equation}
     \ddot q _{\lambda} + \omega _{\lambda}^2q_{\lambda} =
     {4\pi \over c}\int _V \vec J(\vec r,t) \vec A _{\lambda}(\vec r)dV.
     \end{equation} 

The expression (4) is the equation of the oscillator of unit mass
excited by a force $f(t) = (4\pi/c)\int _V \vec J(\vec r,t) \vec A
_{\lambda}(\vec r)dV$.  It describes the excitation of both enclosed
and open resonators \cite{heitler} - \cite{lopuhin}. The same
expression for force determine the excitation of waveguides
\cite{jackson}.

The eigenmodes of the rectangular resonators (cavities) were
discovered by J.Jeans in 1905 when he studied the low of thermal
emission. The equations (4) was used later for quantization of the
electromagnetic field in quantum electrodynamics \cite{heitler}.

\section{Emission of electromagnetic radiation by electron beams in
open resonators}

The equation (4) does not take into account the energy losses of the
emitted radiation in the resonator. These losses can be introduced
through the quality of the resonator $Q _{\lambda}$

     \begin{equation}
     \ddot q _{\lambda} + {\omega _{r}\over Q _{\lambda}}\dot q
     _{\lambda}+ \omega _{\lambda}^2q_{\lambda} = {4\pi \over c}\int _V
     \vec J(\vec r,t) \vec A _{\lambda}(\vec r)dV,
     \end{equation} 
where in the case of the open resonator $\omega _r = 2\pi/T$, $T =
2L/c$ is the period of oscillations of the light wavepacket between the
resonator mirrors when it passes along the axis of the resonator
(notice that in general case the frequencies $\omega _{\lambda} =
\omega _{mnq}$ depend on $m,n,q$ and slightly differ from frequencies
$\omega _r q$). Here we have introduced a version of a definition of a
resonator quality connected with the frequency $\omega _r$.  Another
version of a quality is usually connected with the frequency $\omega
_{\lambda}$.  Our definition is more convenient for the case of
free-electron lasers using open resonators.

Using (5) we can derive the expression for the energy balance in the
resonator. For this purpose we can multiply this equation by $\dot q
_{\lambda}$ and integrate over the volume of the resonator. Then we
receive the equation

     \begin{equation}
     {1\over 2}{d\over dt}[\dot q _{\lambda}^2 + \omega _{\lambda}
     ^2 q _{\lambda}^2] + ({\omega _r\over Q _{\lambda}})^2 \dot q
     _{\lambda}^2 = 4\pi \int _V \vec J(\vec r,t) \vec E _{\lambda}
     (\vec r,t)dV.   \end{equation} 

If we take into account that $\vec E_{\lambda} (\vec r,t) = - \dot q
_{\lambda}(t) \cdot \vec A_{\lambda}(\vec r)/c$, $\vec H _{\lambda}
(\vec r,t) = rot \vec A_{\lambda} (\vec r)$, $rot \vec A _{\lambda} =
\omega _{\lambda}\vec A _{\lambda}/c$, $\int |\vec A_{\lambda}(\vec r)|
^2dV = 1$ then the energy of the free electromagnetic field in the
resonator can be presented in the form $\varepsilon _{\lambda} ^{em} =
\int [({|\vec E _{\lambda}|^2 + |\vec H _{\lambda}| ^2) /8\pi}]$ $dV =
[\dot q _{\lambda}^2 + \omega _{\lambda} ^2 q _{\lambda} ^2]/8\pi c^2$
and the equation (7) can be presented in the another form

     \begin{equation} \dot \varepsilon _{\lambda} ^{em} + (\omega _r/Q
     _{\lambda})\varepsilon _{\lambda} ^{em} = \int _V \vec J(\vec r,t)
     \vec E _{\lambda} (\vec r,t)dV.  \end{equation} 

The equation (5) is the pendulum equation with a friction. It determine
the time evolution of the electromagnetic field stored at the
resonator, when the time dependence of the beam current $\vec J(\vec
r,t)$ is given. The amplitude $q _{\lambda}(\vec r,t)$ according to (5)
is determined by the coefficient of expansion of the given current into
series of eigenfunctions of the resonator. Notice that the value $\vec
A _{\lambda} [\vec r _e(t)] $ depends on $t$ only through $\vec r
_e(t)$ and the value $\vec E _{\lambda}[t, \vec r _e(t)] = - \dot q
_{\lambda}(t) \cdot \vec A_{ \lambda}[\vec r _e(t)]/c$ depends on $t$
directly through $q _{\lambda}(t)$ and through $\vec r _e(t)$.

In the case of one particle of a charge "e" the beam current density
$\vec J(\vec r,t) = e\vec v(t) \delta [\vec r - \vec r_e(t)]$. In this
case the force $f(t) = e \vec v[\vec r_e(t)]\vec A _{\lambda}[\vec r
_e(t)]$ and the power transferred from the electron beam to the
resonator wave mode $\lambda$ excited in the resonator $P _{\lambda}
(t) = \int _V \vec J(\vec r,t) \vec E _{\lambda} (\vec r,t)dV = e \sum
_i \vec v _{e\,i} (t) \vec E _{\lambda} [(\vec r _{e,\i}(t), t)]$.
Using these expressions of force and power for all electrons "i" of the
beam we can present the equations (5), (7) in the form

     \begin{equation}
     \ddot q _{\lambda} + \omega _{\lambda}^2q_{\lambda} =
     {4\pi e\over c}\sum _{i} \vec v _{e \,i}(t) \vec A _{\lambda}[\vec
     r _{e \,i}(t)],   \end{equation} 

     \begin{equation} 
     \dot \varepsilon _{\lambda} ^{em} + (\omega _r/Q
     _{\lambda})\varepsilon _{\lambda} ^{em} = e \sum _i \vec v _{e\,i}
     (t) \vec E _{\lambda} [(\vec r _{e \,i}(t), t)].
     \end{equation}

It follows from (5), (7) and (8), (9) that transverse resonator modes
are excited only in the case when the force $f(t) \ne 0$ and the power
$P _{\lambda}(t)\ne 0$ that is when the particle trajectory passes
through the regions where the corresponding resonator modes have large
intensities and when the particle velocity has transverse and/or
longitudinal components directed along the direction of the electric
field strength. Open resonators on the level with enclosed ones have
modes with longitudinal components of electric field strength (see
Appendix). It means that open resonators can be excited even in the
case when the particle trajectory have no transverse components and its
velocity is directed along the axis of the resonator\footnote{In
this case the transition radiation is emitted by particles when they
pass the walls of a resonator. The electromagnetic radiation will be
emitted in the form of thin spherical layers at the first and second
resonator mirrors \cite{shib}. It will be reflected then repeatedly by
resonator mirrors. The expansion of the electromagnetic fields of the
spherical layers will be described by the series (2).}. Using external
fields of a single bending magnet can increase the power of the
generated radiation. Both in the case of lack of a banding magnet and
presence of one bending magnet the broadband radiation is emitted.  The
experiment confirms this observations \cite{shibata}.  Using external
fields of undulators and beams bunched at frequencies of the emitted
radiation can lead to emission of rather monochromatic radiation.

In the simplest case when the beam current density $\vec J(\vec r,t)
$ is a periodic function of time then the force can be expanded in the
series $f(t) = \int _V \vec J(\vec r,t) \vec A _{\lambda}(\vec r)dV =
\sum _{\nu = -\infty} ^{\infty}f _{\lambda \nu}\exp[i(\nu \omega _bt -
\varphi _{\lambda \nu})]$, where $\omega _b = 2\pi / T _b$ and $T _b$
are a period and frequency of the current density oscillation
accordingly, $f _{\lambda \nu} = (1/T _b) \int _{-T_b/2} ^{T_b/2}
f(t)\exp(i\nu \omega _b t)dt$, are the known coefficients, $\varphi
_{\lambda \nu}$ phase.  The value $f_{\lambda\, - \nu} = f ^{*}
_{\lambda \,\nu}$, where $f ^{*} _{\lambda \nu}$ is the complex
conjugate of $f _{\lambda \nu}$. The solution of the equation (5) for
the case of the established oscillations ($t \gg Q _{\lambda}T _b$) is

     \begin{equation}
     q _{\lambda}(t) = \sum _{\nu = 1} ^{\infty}A _{\lambda \nu
     }\exp[i(\nu \omega _bt - \theta _{\lambda \nu})],
     \end{equation} 
     where
     $$A _{\lambda \nu} = {f _{\lambda \nu}\over
\sqrt{(\omega ^2_{\lambda} - \nu ^2\omega _b^2)^2 + (\nu \omega
_{r}\omega _b/Q_{\lambda})^2}},$$

     $$\theta _{\lambda \nu} = \varphi _{\lambda \nu} + arctg
{\nu\omega _{r}\omega _b\over Q_{\lambda}(\omega ^2_{\lambda} -
\nu ^2\omega _b^2)}.$$

It follows from the equation (10) that the maximum of the amplitude of
the vector potential $A _{\lambda \nu} = Q _{\lambda}f _{\lambda \nu}
/\omega  _{r}\omega ^2 _{\lambda}$ takes place at resonance $\nu \omega
_b = \omega _{\lambda} = \omega _{mnq} \simeq \omega _r q$. Notice that
all modes $\lambda$ are excited at the same frequency $\omega _b$ of
the oscillator. In general case $\omega _b \ne \omega _{\lambda}$.

The equation (10) is the first order linear equation of the energy
change in the resonator excited by the electron beam. It follows from
this equation that after switching off the beam current at some moment
$t _0$ ($\vec J(\vec r,t)|_{t>t_0} = 0$) the energy in the resonator
will be changed by the law $\varepsilon _{\lambda} ^{em} = \varepsilon
_{\lambda ,/0} ^{em}\exp[-(t - t _0)/ \tau]$, where $\tau = Q
_{\lambda} /\omega _r$, $\varepsilon _{\lambda \,0} ^{em} =
\varepsilon _{\lambda } ^{em}|_{t = t _0}$. On the contrary after
switching on the beam current at some moment $t _0$ the energy in the
resonator will be changed by the law $\varepsilon _{\lambda} ^{em} =
\varepsilon _{\lambda m} ^{em}(1 - \exp[-(t - t _0))/ \tau]$, where the
energy of the electromagnetic field in the resonator $\varepsilon
_{\lambda m} ^{em}$ is determined by the parameters of the resonator and
the beam.

The considered example describes the emission of an oscillator or a
system of oscillators which are in phase with the excited mode and have
zero average velocity (trajectory has a form $r _e = r _{e0} + \vec
\imath a_0\cos \omega _0 t$, where $\vec \imath$ is the unite vector
directed along the axis $x$). More complicated examples of trajectories
of particles using for excitation of resonators by electron beams can
be considered (the arc of circle, sine- or helical-like trajectories in
bending magnets and undulators).

\section{Vector TEM modes of open resonators}

The theory of high quality open resonators does not differ from
enclosed ones. But eigenmodes of open resonators have some unique
features. The spectrum of the open resonators is rarefied, the
operating mode spectrum has maximum selectivity. The dimensions of open
resonators are much higher then the excited wavelengths and the
dimensions of the enclosed resonators are of the order of excited
wavelengths. The quality of open resonators at the same wavelengths is
higher then enclosed ones.

There are some methods of calculation of TEM modes in open resonators.
Usually scalar wave equations are investigated \cite{maitland},
\cite{svelto}. There is a small information in technical publications
about distribution of vectors of the electric and magnetic field
strengths in such resonators. In this section we search some
distributions. In the Appendix the foundations of the
excitation of resonators by electron beams are presented .

We will present the result for the Cartesian coordinates. In this case
the solution of the scalar wave equation (24) (see Appendix) has a form
\cite{oraevskiy}

        $$V_{mn}(x,y,z) = {C\over\sqrt{w _x(z)w _y(z)}}
        H _m\left({\sqrt 2x\over w _x(z)}\right)H _n\left({\sqrt
        2y\over w _y(z)}\right) \cdot$$
        \begin{equation} 
        \exp \left\{ {ik\over 2}\left({x ^2\over
        q _x(z)} + {y ^2\over q _y(z)}\right) - i(m + {1\over 2}) arctg
        {\lambda z\over \pi w _{0x}^2} - i(n + {1\over 2}) arctg
        {\lambda z\over \pi w _{0y}^2} \right\} \end{equation}
and for the cylindrical coordinates

        \begin{equation} 
        V(r,\phi,z) = C\left({r\over w(z)}\right) ^m{\sin m\phi
        \choose \cos m\phi}L ^m _n\left({2r ^2\over w
        ^2(z)}\right)\exp \left\{ {ikr^2\over q(z)} - i(m + 2n +1)
        arctg {\lambda z\over \pi w_0^2}\right\}w(z) ^{-1},
        \end{equation}
where $H _m$, $H _n$ are the Hermittian polynomials, $L ^m_n$ the
Lagerian polynomials, $\lambda = 2\pi c/\omega$ is the wavelength, C =
constant,

       $${1\over q(z)} = {1\over R(z)} + {i\lambda \over \pi w
       ^2(z)}, \hskip 5mm  R(z) = z\left[1 + \left({\pi w _0^2\over
       \lambda z}\right) ^2\right], \hskip 5mm  w ^2(z) =
       w_0^2\left[1 + \left({\lambda z\over \pi w _0^2}\right)
       ^2\right].$$

In (11), (12) $R(z)$ is the radius of the wave front of Gaussian beam,
$w (z)$ the radius of the beam, $w _0(z)$ the radius of the waist of
the beam.

At $m = n = 0$ we have the main mode of the Gaussian beam. If $w _{0x}
= w _{0y} = w _0$ then the main modes for the Cartesian and cylindrical
coordinates are the same

        \begin{equation} 
        U(x,y,z) = {C\over w(z)}\exp \left\{ - {x ^2 + y^2\over w
        ^2(z)} + {ik\over 2}{x ^2 + y ^2\over R(z)} - i\,
        arctg {\lambda z\over \pi w _0^2} \right\}\exp^{i(kz -
        \omega t)}.  \end{equation}

We have the solutions (11), (12) of the scalar wave equation (24) for
the space limited beam. Now we can find vectors of the electric and
magnetic field strengths using the expressions (23) and possible ways
of construction of Hertz vectors. Let us suppose the next compositions
with the electric Hertz vector assuming that magnetic Hertz vector is
zero:\\

1) $\Pi ^e_x = U(x,y,z)$, $\Pi ^e_y = \Pi ^e_z = 0$. \\

2) $\Pi ^e_x = 0$, $\Pi ^e_y = U(x,y,z)$, $\Pi ^e_z = 0$. \\

3) $\Pi ^e_x = 0$, $\Pi ^e_y = 0$, $\Pi ^e_z = U(x,y,z)$. \\

In the first case \\

$div \vec \Pi = \partial \Pi _x/\partial x = \partial V/\partial x \exp
[i(kz -\omega t)]$, $(rot \vec \Pi)_x = 0$,

$(rot \vec \Pi)_y = (\partial V/\partial z + ikV)\exp [i(kz -\omega
t)]$, $(rot \vec \Pi) _z = - (\partial V/\partial y)\exp [i(kz -\omega
t)]$\\

and\\

$E _x^1 = \partial ^2V/\partial x^2 + k^2V$, $E _y^1 = \partial
^2V/\partial x \partial y$,

$E _z^1 = \partial ^2V/\partial x \partial z + ik \partial V/\partial
x$, $H _x^1 = 0$, $H _y^1 = ik\partial V/\partial z - k^2V$, $H _z^1 =
ik\partial V/\partial y$.    \\

The upper superscript shows the first composition of the Hertz vector.
A common multiple $\exp [i(kz -\omega t)]$ for all field components is
omitted.

The values $\partial ^2V/\partial x_i \partial x_k \ll k\partial
V/\partial x_i \ll k^2V$. That is why in this case $E _x^1 \gg E _y^1,
E _z^1$, $ H _y^1 \gg H _z^1$.\\

The second case does not differ from the first one. It is necessary to
substitute variable $x$ by $y$ and vise versa.\\

In the third case \\

$div \vec \Pi = \partial \Pi _x/\partial z = \partial V/\partial x\exp
[i(kz -\omega t)]$, $(rot \vec \Pi)_x = \partial V/\partial y\exp [i(kz
-\omega t)]$,

$(rot \vec \Pi)_y = - \partial V/\partial x\exp [i(kz -\omega t)]$,
$(rot \vec \Pi) _z = 0$\\

and\\

$E _x^3 = \partial ^2V/\partial x \partial z + ik\partial V/\partial
x$, $E _y^3 = \partial ^2V/\partial z \partial y + \partial V/\partial
y$,

$E _z^3 = 2ik \partial V/\partial z$, $H _x^3 = ik\partial V/\partial
y$, $H _y^3 = - ik\partial V/\partial x$, $H _z^3 = 0$.  \\

It follows that in the case of the main mode the electric and magnetic
field strengths corresponding to the electric Hertz vector have
components:

        $$E^1_x = k^2U(x,y,z), \hskip 5mm E^1_y \simeq 0, \hskip 5mm
        E^1_z = 2ikx\left[{1\over w^2(z)} + {ik\over R(z)} \right]
        U(x,y,z),$$

        $$H^1_x \simeq 0, \hskip 5mm H^1_y = - k^2U(x,y,z),\hskip 5mm
        H_z^1 = 2iky \left[{1\over w^2(z)} + {ik\over R(z)} \right]
        U(x,y,z)$$

        $$E^2_x = \simeq 0, \hskip 5mm E^2_y = k^2U(x,y,z), \hskip 5mm
        E^2_z = 2iky\left[{1\over w^2(z)} + {ik\over R(z)} \right]
        U(x,y,z),$$

        $$H^2_x = - k^2U(x,y,z), \hskip 5mm H^2_y \simeq 0,\hskip 5mm
        H_z^2 = 2ikx \left[{1\over w^2(z)} + {ik\over R(z)} \right]
        U(x,y,z)$$

        \begin{equation} 
        E^3_x = 2ikx\left[{1\over w^2(z)} + {ik\over
        R(z)}\right]U(x,y,z), \hskip 5mm E^3_y = - 2iky\left[{1\over
        w^2(z)} + {ik\over R(z)}\right]U(x,y,z),
        \end{equation}

        $$E^3_z = 2ik\left\{{4\lambda (x^2 + y^2)z\over (\pi w
        _0)^2w^3(z)} + {ik(x^2 + y^2)\over 2R^2(z)}\left[1 - {\left(\pi
        w_0^2\over \lambda z\right)^2} \right] - {i\lambda \over \pi
        w_0 ^2 \left[1 + \left({\lambda z\over \pi w_0^2} \right)
        ^2\right]} \right\}U(x,y,z),$$

        $$H^3_x = 2iky\left[{1\over w^2(z)} + {ik\over R(z)}\right]
        U(x,y,z), \hskip 5mm H^3_y = - 2iky\left[{1\over w^2(z)} +
        {ik\over R(z)}\right] U(x,y,z),\hskip 5mm H_z^3 = 0.$$

The electric and magnetic field strengths received from magnetic
Hertz vector can be received from the fields (14) as well. For this
purpose we can take the vector of the electric field strength received
from magnetic Hertz vector equal to the negative value of the magnetic
field strength received from the electric Hertz vector $\vec E ^{'} \to
- \vec H$ and by analogy we can take $\vec H ^{'} \to  \vec E$.

The general solution for the electromagnetic field strength of the main
mode of Gaussian beam $TEM_{00}$ can be presented in the form

        $$\vec E = c_1\vec E^1 + c_2\vec E^2 + c_3\vec E^3 - c_4\vec H^1
        - c_5\vec H^2 - c_6\vec H^3, $$

        \begin{equation} 
        \vec H = c _1\vec H^1 + c_2\vec H^2 + c_3\vec H^3 + c_4\vec E^1
        + c_5\vec E^2 + c_6\vec E^3,
        \end{equation}
where $c_i$ are the arbitrary coefficients determined by the conditions
of excitation of the mode by the electron beam. Waves determined by the
only coefficient $c_i$ (when another ones are equal to zero) can be
excited independently.

Higher modes in the open resonator will be described by the expressions
(15) and by the expressions similar to (14) for the electromagnetic
field strengths of the main mode. They will form orthogonal and
full set of fundamental waves. The arbitrary wave may be expanded into
these waves. Of cause, real electric and magnetic field strengths are
determined by the real part of the expression (15).

In the open resonators the same Gaussian beams are excited. They
propagate between mirrors both in $z$ and in $-z$ directions. However
the resonators will be excited on discrete set of eigenfrequences
(wavelengths) \cite{oraevskiy}.

We can see that according to (14) all considered waves $\vec E ^{i},
\vec H ^i$ are transverse. At the same time they have longitudinal
components. This is the general property of the convergent and
divergent waves \cite{vainstein}, \cite{oraevskiy}. Such waves have
longitudinal components which permit the lines of the electric and
magnetic field strengths to be closed.

The fields $\vec E ^{1}, \vec H ^1$ describe an electromagnetic wave
with one direction of polarization and the fields $\vec E ^{2}, \vec H
^2$ with another one. They have high transverse components of the
electric and magnetic field strengths and zero longitudinal components
on the axis $z$.

Electromagnetic fields $\vec E ^{3}, \vec H ^{3}$ are a new kind of
fields. They have zero transverse components of the electric and
magnetic field strengths and high value longitudinal component of the
electric field strength at the axis $z$ (similar to the wave $E_{01}$
at the axial region of the cylindrical waveguide). It means that in
this case the lines of the  electric and magnetic field strengths are
closed in the directions both at the central part of the beam
propagation that is near to the axis $z$ and far from the axis that is
near to the region of theirs envelopes (caustics)\footnote{Notice that
usually the divergent waves with high directivity emitted by antennas
are described and drawn by the lines of the electric and magnetic field
strengths which are closed in the directions far from the axis of the
beam propagation near to the region of theirs envelops.}.

Usually the scalar functions $V(x,y,z)$ or $U(x,y,z) = V(x,y,z) \exp
[i(kz - \omega t)]$ are used when the modes in open resonators are
investigated \cite{vainstein}, \cite{maitland}, \cite{svelto}. It
was supposed that the waves are transverse ones and the values of the
electromagnetic field strengths are distributed near the same way as
the values of the scalar functions. At that some features like the
existence of the wave $\vec E ^{3}, \vec H ^{3}$ were hidden. Such
waves have longitudinal components of the electric field strength and
hence can be excited through the transition radiation emitted on the
inner sides of the resonator walls by an electron homogeneously moving
along the axis $z$.  Such excitation was observed in the experiments
published in \cite{shibata}.

\section{Conclusion}

Open resonators permit an effective generation of broadband radiation
at the main and/or other transverse modes under conditions when many
longitudinal modes are excited. The longitudinal modes are limited in
the longwavelength region by the diffraction losses and in the short
wavelength region by the longitudinal electron beam dimensions
(coherence conditions). Open resonators can be excited in the case when
the external fields in the resonator are absent and the particle
trajectory is directed along the axis of the resonator. Using external
fields of a single bending magnet can increase the power of the
generated radiation \cite{shibata}.

\vskip 10mm
\begin{center}
\renewcommand{\appendixname}{\large \bf Appendix}\appendixname
\end{center}

\appendix

Generation and propagation of electromagnetic waves in vacuum is
described by Maxwell equations (1). We noticed above that these
equations are a set of eight equations for six independent components
of the electric and magnetic fields. Only four components of the
electromagnetic field are independent. These equations added with
initial and boundary conditions describe all processes in
electrodynamics.

There is no general solution of the system of Maxwell equations with
boundary conditions similar to the Lienard-Viechert solution for the
fields produced by charged particles moving along some trajectories at
a given low in free space. It means that private problems must be
solved separately for every concrete case. At that when the boundary
conditions exist, interactions of particles with surrounding media and
intrabeam interactions of particles are essential then the beam density
and beam current can not be given and the dynamical Lorentz equations
must be added. Below we will consider the case when the beam density
and the density of the beam current (particle trajectories) are given.

One of the possible simplifications of the solution of the Maxwell
equations is to transform linear Maxwell equations to the equations of
the second order relative to the field strengths or potentials.

First of all the Maxwell equations can be transformed to the equations
separately for the electric and magnetic fields. For this purpose we
can differentiate equation (1.b) with respect to $t$, use equation
(1.c) and employ the vector identity $rot\,rot\vec{F} =
grad\,div\,\vec{F} - \Delta \vec{F}$, where $\Delta $ is the Laplacian
operator. Such a way we will receive the equation for the electric field
strength and then by analogy we will receive the equation for the
magnetic field strength. They are

        \begin{equation} 
        \Box \vec{E}=\frac{4\pi}{c^2} \dot{\vec{J}} + 4\pi
        \,grad\,\rho, \hskip 3mm (a)  \hskip 5mm \Box \vec{H} = -
        \frac{4\pi}crot\,\vec{J}.   \hskip 3mm (b) \end{equation}
where $\Box = \Delta - \partial ^2/c^2\partial t^2$ is the d'Alembertian
operator, $\dot{\vec{J}} = \partial{\vec J}/\partial t$.

The equations (16) are the nonhomogeneous linear equations of the second
order. We must add the equations (1.a), (1.d) to the system of the
equations (16). It means that we have again a system of two vector and
two scalar equations (in components they are eight equations) for six
unknown components of the electric and magnetic field strengths $E_i$,
$H_i$.

The divergence of the equation (16.a) leads to a more general continuity
equation $(\partial/\partial t)(\partial \rho/$ $\partial t$ $ + div
{\vec{J}}) = 0$ which is valid when the continuity equation $\partial
\rho/\partial t + div {\vec{J}} = 0$ is valid.

The solution of the Maxwell equations will be the solution of these
second order equations.  The second order equations are another
equations.  Strictly speaking they are not equivalent to Maxwell
equations. We must check theirs solutions by substituting these
solutions into the linear Maxwell equations to reject unnecessary
solutions. This is very difficult problem even for simple cases. A way
out can be found by introducing of electromagnetic field potentials.
The vector potential $\vec A$ and scalar potential $\varphi$ are
introduced by the equations $\vec H = rot \vec A$, $\vec E = - grad
\varphi - (1/c)(\partial \vec A/\partial t)$. In this case both from
Maxwell equations and from the equations (16) it follows the equations
for vector and scalar potentials

        \begin{equation} 
        \Box \vec{A}= - \frac{4\pi}{c} {\vec{J}}
        \hskip 3mm (a),  \hskip 5mm \Box \varphi = -
        {4\pi}\rho  \hskip 3mm (b) \end{equation}
and additional condition coupling the potentials (Lorentz gauge)

        \begin{equation} 
        div \vec{A}= - \frac{1}{c} \frac{\partial \varphi}{\partial t}.
        \end{equation}

It is convenient to use the electric and magnetic Hertz vectors as
well. They permit to simplify the solutions of the problem of
propagation of waves in resonators and free space which is described by
the homogeneous wave equations ($\rho = 0$, $ \vec J = 0$). Both the
electric and magnetic Hertz vectors $\vec \Pi ^e$, $\vec \Pi ^m$ are
introduced by the same expressions

        \begin{equation} 
        \vec{A}= \frac{1}{c} \frac{\partial \vec \Pi ^{e/m}}{\partial
        t}; \hskip 8mm \varphi = - div \vec \Pi ^{e/m}.  \end{equation}

Such way defined potentials $\vec A$ and $\varphi$ will satisfy the
equation (11) simultaneously.

Different superscripts $e/m$ in this case are used on the stage of
introduction of the connection between electric and magnetic field
strengths through Hertz vectors. The electric field strength can be
expressed through the electric and magnetic Hertz vector by the
equations

        \begin{equation} 
        \vec{E} = grad\, div \vec \Pi ^e - {1\over c ^2}\frac {\partial
        ^2 \vec \Pi ^e}{\partial t^2}; \hskip 8mm \vec H = {1\over c}
        {\partial \over \partial t} rot \vec \Pi ^e,  \end{equation}

        \begin{equation} 
        \vec{E} = - {1\over c}{\partial\over \partial t} rot \vec \Pi
        ^m, \hskip 8mm \vec H = grad\, div \vec \Pi ^m - {1\over c
        ^2}\frac {\partial ^2 \vec \Pi ^m}{\partial t^2}.
        \end{equation}

These manipulations are valid because of both definitions (20) and (21)
satisfy Maxwell equations (1) and equations (16). This is because of
homogeneous wave equations for electromagnetic fields
\vskip -3mm
        \begin{equation} 
        \Box \vec{F} = 0 \end{equation}
are symmetric relative to fields $\vec E$, $\vec H$ ($\vec F = \vec
E, \vec H$). If $\vec E$ and $\vec H$ are some solutions of the
homogeneous Maxwell equations (1b), (1c) then vectors $\vec E ^{'} = -
\vec H$ and $\vec H ^{'} = \vec E$ will satisfy these and
another Maxwell equations as well.

In general case the problem may be reduced to solving of wave equation
if potentials $\vec \Pi ^e$, $\vec \Pi ^m$ will be introduced
simultaneously in the form \cite{oraevskiy}

        \begin{equation} 
        \vec{E} = grad\, div \Pi ^e - {1\over c ^2}\frac {\partial
        ^2 \vec \Pi ^e}{\partial t^2} - {1\over c}{\partial\over
        \partial t} rot \vec \Pi ^m, \hskip 8mm \vec H = {1\over c}
        {\partial \over \partial t} rot \vec \Pi ^e + grad\, div \vec
        \Pi ^m - {1\over c ^2}\frac {\partial ^2 \vec \Pi ^m}{\partial
        t^2}.  \end{equation}

We can be convinced that $\vec E$ and $\vec H$ described by (23)
fulfil to Maxwell equations at $\rho = \vec J = 0$ when vectors
$\vec \Pi ^e$ and $\vec \Pi ^m$ fulfil the wave equation (22) with
replaced $\vec F \to $ on $\vec \Pi ^e$ and $\vec \Pi ^m$.

Equation (22) is valid for each component of vectors $\vec \Pi ^e$ and
$\vec \Pi ^m$. That is why it is possible to use scalar wave equation

        \begin{equation} 
        \Box U = 0 \end{equation}
and identify its solution $U$ with one of components of vectors $\vec
\Pi ^e$ or $\vec \Pi ^m$ and the rest components of these vectors
equate to zero (say we can take $\vec \Pi ^e = \vec e _x \cdot 0 + \vec
e _y \cdot 0 + \vec e _z \cdot U,$\hskip 2mm $ \vec \Pi ^m = 0$).
Substituting the constructed such a way vector with one component in
(16) we will find the electromagnetic field strengths $\vec E$, $\vec
H$ which satisfy the Maxwell and wave equations. Then we can identify
the same solution with another component of the Hertz vector, equate
the rest components to zero and calculate another electromagnetic field
strengths $\vec E$, $\vec H$ which satisfy the Maxwell and wave
equations as well. After we will go through all compositions with
components then we will have a set of six different solutions for field
strengths $\vec E$, $\vec H$.  These solutions will be six
electromagnetic waves with different structures. Sum of these solutions
with some coefficients will be a solution of the Maxwell equations as
well. This will be algorithm of electromagnetic field determination
through Hertz vector.

Equation (24) has many different solutions. We must find such solutions
which will correspond to the problem under consideration to a
considerable extent. Below we will deal with monochromatic light beams
of the limited diameter related with resonator modes. In general case
such beams can be written in the form

        \begin{equation} 
        U(x,y,z) = V(x,y,z)e^{i(kz -\omega t)}
        \end{equation}
where $V(x,y,z)$ is a function of coordinate slowly varying in
comparison with $\exp{i(kz -\omega t)}$. A complex form of values will
be used for computations and then we will proceed to a real part of the
form.

Substituting (25) into (24) and taking into account the slow variation
of $V(x,y,z)$ compared with $\exp{i(kz -\omega t)}$ that is the
condition $|\partial^2 V/\partial z ^2| \ll 2k|\partial V/\partial z|$
and the condition $k = \omega /c$ we will receive the equation

        \begin{equation} 
        i{\partial V\over \partial z} + {1\over 2k}({\partial ^2 V\over
        \partial x^2} + {\partial ^2 V\over \partial y^2}) = 0
        \end{equation}
which describes a space limited beam.

In the general case the limited in the transverse direction wave
propagating in free space or in a resonator have rather complicated
structure. That is why it is desirable to find full, orthogonal set
of fundamental waves with the well known feature of propagation. Then
an arbitrary wave may be expanded into series of these waves. Different
series of fundamental waves can be found for this problem and the
arbitrary wave can be expanded into one or another series. The method of
separation of variables is used to solve the wave equation. For
example, in the Cartesian coordinates $V(x,y,z) = X(x,y,z)\cdot
Y(x,y,z)$ and in the cylindrical coordinates $V(x,y,z) = G(u)\Phi
(\varphi)\exp [ikr ^2/2q(z)]\cdot \exp[iS(z)]$, where $r$ and $\varphi$
are cylindrical coordinates on a plane transverse to $z$, $u =
r/w(z)$.  These solutions are considered in \cite{oraevskiy}.

\end{document}